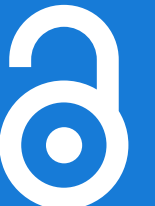

# OGRePy: An Object-Oriented General Relativity Package for Python

BARAK SHOSHANY

## ABSTRACT

OGRePy is a modern, open-source Python package designed to perform symbolic tensor calculations, with a particular focus on applications in general relativity. Built on an object-oriented architecture, OGRePy encapsulates tensors, metrics, and coordinate systems as self-contained objects, automatically handling raising and lowering of indices, coordinate transformations, contractions, partial or covariant derivatives, and all tensor operations. By leveraging the capabilities of SymPy and Jupyter Notebook, OGRePy provides a robust, user-friendly environment that facilitates both research and teaching in general relativity and differential geometry. This Python package reproduces the functionality of the popular Mathematica package OGRe, while greatly improving upon it by making use of Python's native object-oriented syntax. In this paper, we describe OGRePy's design and implementation, and discuss its potential for reuse across research and education in mathematics and physics.

**CORRESPONDING AUTHOR:**
**Barak Shoshany**
Brock University, Canada
bshoshany@brocku.ca

## (1) OVERVIEW

### INTRODUCTION

Tensor calculus is a cornerstone of modern theoretical physics, particularly in the study of general relativity. However, the intricate bookkeeping required when manipulating tensorial quantities—such as raising and lowering indices, performing contractions, or transforming between coordinate systems—often makes manual calculations tedious and error-prone. OGRePy addresses these challenges by providing an object-oriented framework that encapsulates all tensor operations within self-contained Python objects. Based on the popular OGRe package [1], originally written for Wolfram Mathematica in 2021, OGRePy not only replicates its functionality but also enables a simpler and more intuitive syntax by taking full advantage of Python's object-oriented programming capabilities.

At its core, OGRePy represents tensors, metrics, and coordinate systems as distinct objects that automatically manage their components in different representations. A tensor is initially created in one coordinate system and index configuration, but transforms into different coordinates and/or index configurations on the fly as required by the specific operations being performed on it, ensuring that the underlying mathematical structure is consistently preserved. This design eliminates user error and simplifies tensor manipulations of arbitrary complexity.

### IMPLEMENTATION AND ARCHITECTURE

OGRePy is implemented in Python (version 3.12 and above) and leverages the symbolic manipulation capabilities of SymPy [2], as well as the interactive user interface of Jupyter notebooks [3, 4]. Its architecture is built around three principal object classes:

1. A `Coordinates` object represents a symbolic coordinate system on a manifold. By registering explicit transformation rules between different sets of coordinates, OGRePy can seamlessly switch representations behind the scenes.
2. A `Metric` object represents the metric tensor for a manifold. This class provides many methods for calculating related quantities, such as the line and volume elements, the Christoffel symbols, various curvature tensors such as the Riemann, Ricci, and Einstein tensors and the Ricci and Kretschmann scalars, and the geodesic equations in various representations.
3. A `Tensor` object represents a tensor of arbitrary rank on a manifold, including scalars, vectors, and higher-rank tensors. Upon construction, the user provides the tensor's initial components in a particular coordinate system and index configuration. Other representations of the tensor are calculated on the fly as needed by raising and lowering indices with the metric, or changing coordinates through user-defined transformations. Internally, the tensor components are stored in a dictionary that maps a tuple (`indices, coords`) to a SymPy `Array` object containing the components in that representation; this is initially populated with the representation given by the user, and new representations are automatically added as needed.

In addition, the package utilizes many classes derived from the main `Tensor` class behind the scenes, for special cases such as curvature tensors and geodesic equations.

The object-oriented approach has many advantages, including:

- Encapsulation: Each tensor object contains its components in all representations (at least those that have been used so far). This ensures that different representations of the same abstract tensor are not confused with each other, and that the user does not need to calculate any representations manually.
- Safety: Users are shielded from accidental "illegal" operations such as adding tensors of different rank, mixing tensors from different metrics, or contracting more than 2 indices. OGRePy will automatically display informative error messages for such violations.
- Caching and Efficiency: Once OGRePy computes a particular representation of a tensor, it remembers the result, avoiding the need for repeated computation in later steps. All objects derived from metrics, such as Christoffel symbols, curvature tensors, and geodesic equations are also cached, for significantly increased performance.

OGRePy integrates seamlessly with Jupyter Notebook, where its `show()` and `list()` methods automatically render tensors of any rank in a human-readable format (using TeX behind the scenes). This combination of symbolic precision and interactivity makes OGRePy an excellent tool both for research and pedagogy.

### QUALITY CONTROL

All aspects of OGRePy's codebase are tested with each release, to ensure that any modifications do not introduce bugs or regressions, and that new features work as expected. The package includes comprehensive run-time error checking and validation routines that prevent illegal tensor operations, as mentioned above.

In addition, the comprehensive use of type checks at runtime enables OGRePy to detect programming errors, assisting both novice and experienced users in debugging their code. For example, passing an argument of the wrong type to the constructor of any of the classes listed above will result in an error message instead of creating an invalid object.

OGRePy was written from scratch with performance in mind, utilizing fast algorithms, efficient memory use, caching, and other techniques. However, the speed of symbolic manipulations is fundamentally limited by the performance of SymPy. Mathematica (written in C/C++) has much faster symbolic capabilities than SymPy (written in pure Python), so OGRe often performs calculations much faster than OGRePy. In addition, OGRe supports multi-core parallelization by launching multiple Mathematica kernels, while OGRePy does not, due to Python's global interpreter lock.

To illustrate, we conducted internal benchmarks on an Intel Core i9-13900K CPU. Calculating the Einstein tensor of the FLRW metric took ~270 ms in OGRe and ~2400 ms in OGRePy (taking the average of 5 runs), indicating a 9x speedup for OGRe. Parallelization was not enabled in OGRe, so only a single core was used in both cases. Since such calculations only need to be performed once and are then cached, we found that this performance penalty has little effect in practice, and we expect that future versions of SymPy will have improved performance.

Each class and method used in this package is accompanied by a detailed docstring, which can be displayed automatically when editing the code in an IDE such as Visual Studio Code. Extensive documentation, with numerous usage examples, is available in multiple formats (Markdown, Jupyter Notebook, HTML, and PDF) on the GitHub page. In addition, when the package is imported into a Jupyter notebook, it displays links to the documentation in various formats on the screen for easy access.

## (2) AVAILABILITY

### OPERATING SYSTEM

OGRePy is platform-independent and runs on any operating system supporting Python 3.12 and above, including all recent versions of Windows, Linux, and macOS.

### PROGRAMMING LANGUAGE

Python (v3.12+).

### ADDITIONAL SYSTEM REQUIREMENTS

None.

### DEPENDENCIES

OGRePy requires ipykernel (v6.29+) and sympy (v1.13+). These dependencies are automatically installed when installing OGRePy from PyPI using `pip install OGRePy`. The notebook interface of Visual Studio Code (v1.97+) is the officially recommended way to use the package, due to helpful features such as IntelliSense, tooltips, and type checking, but it can also be used with JupyterLab.

### LIST OF CONTRIBUTORS

None.

### SOFTWARE LOCATION

**Archive**

***Name:*** Zenodo
***Persistent identifier:*** doi:10.5281/zenodo.14880011
***Licence:*** MIT
***Publisher:*** Barak Shoshany
***Version published:*** v1.3.0
***Date published:*** 16/02/2025

**Code repository**

***Name:*** GitHub
***Identifier:*** https://github.com/bshoshany/OGRePy
***Licence:*** MIT
***Date published:*** 16/02/2025

### LANGUAGE

English

## (3) REUSE POTENTIAL

OGRePy's flexibility and comprehensive feature set make it broadly reusable in all fields requiring symbolic manipulation of tensors. Its intended use case is for research in general relativity and differential geometry, including fields such as gravitational physics, high-energy physics, quantum gravity, astrophysics, and cosmology. In the months since its release, it has already been successfully used in multiple research projects by the author and his students. In addition, the package can be used for teaching advanced undergraduate or graduate courses in general relativity and differential geometry.

OGRePy's source code is freely available on GitHub via the permissive MIT license, and forks and contributions are encouraged. Prospective developers are invited to open issues and submit pull requests on GitHub. Support mechanisms for users include GitHub issues, which can be used for bug reports, feature requests, and technical support, as well as simply emailing the author at bshoshany@brocku.ca.

Users are invited to notify the author about any research or teaching projects utilizing this package. Such projects will be featured in the GitHub repository, including links to associated publications and/or code if available.

## DATA ACCESSIBILITY STATEMENT

There is no data associated with this package. The source code can be found in the GitHub repository.

## ACKNOWLEDGEMENTS

I would like to thank my student Jared Wogan, whose undergraduate research project, a preliminary Python port of my Mathematica package OGRe, motivated and inspired me to eventually write my own port, OGRePy.

## FUNDING STATEMENT

I acknowledge the support of the Natural Sciences and Engineering Research Council of Canada (NSERC), RGPIN-2024-04063.

## COMPETING INTERESTS

The author has no competing interests to declare.

## AUTHOR AFFILIATIONS

**Barak Shoshany** orcid.org/0000-0003-2222-127X
Brock University, Canada

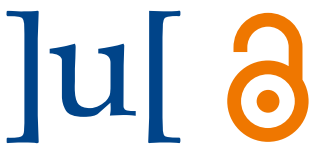